\documentclass[11pt]{article}
\usepackage{graphicx}
\usepackage{epsf}% Include figure files
\usepackage{epsfig}
\usepackage{amsfonts}

\begin{document}
\title{New physics, the cosmic ray spectrum knee, and $pp$ cross section measurements}
\author{Aparna Dixit$^1$, Pankaj Jain$^2$, Douglas W. McKay$^3$, and
Parama Mukherjee$^4$}

\maketitle
\begin{center}
{\small 1) Physics Department, PSIT, Kanpur, India\\
2) Department of Physics, IIT Kanpur, Kanpur 208016, India\\
3) Department of Physics \& Astronomy, University of Kansas,\\ Lawrence, KS - 66045, USA\\
4) School of Marine and Atmospheric Sciences,
Stony Brook University,\\ Stony Brook, NY 11794, U.S.A}
\end{center}

\begin{abstract}
We explore the possibility that a new physics interaction can provide an explanation for the knee just above $10^6$ GeV in the cosmic ray spectrum.  
We model the new physics modifications to the total proton-proton cross section with an incoherent term that allows for missing energy above the scale of new physics. 
We add the constraint that the new physics must also be consistent with published $pp$ cross section measurements, using cosmic ray observations, an order of magnitude and more above the knee. We find that the rise in cross section required at energies above the knee is radical.   
The increase in cross section suggests that it may be more appropriate to treat the scattering process in the black disc limit at such high energies.  In this case there may be no clean separation between the standard model and new physics contributions to the total cross section. We model the missing energy in this limit and find a good fit to the Tibet III cosmic ray flux data. We comment on testing the new physics proposal for the cosmic ray knee at the Large Hadron Collider. 
\end{abstract}

\section{Introduction}
A power law energy dependence going approximately as $E^{-2.7}$ characterizes the high energy cosmic ray spectrum from 10 GeV to $10^6$ GeV.  In the range $10^6$ GeV to $4\times10^6$ GeV, the energy dependence changes to $E^{-3}$, continuing with this slope to $10^9$ GeV or so.  The downward bend just above $10^6$ GeV is called the ``knee" in the spectrum.
This knee phenomenon in the cosmic ray spectrum \cite{knee_disc}, observed by many experiments over many years \cite{akeno, casa-mia, kaskade, tibet}, and reviewed recently in \cite{beh}, still lacks a convincing explanation. 
It is generally believed to be of astrophysical
origin, caused by leakage of cosmic rays from the galaxy or due to an upper
limit on the energy that cosmic ray particles can attain at supernova remnants,
the galactic acceleration sites \cite{beh}.
The center of mass energy corresponding
to the knee is several TeV in the proton-proton ($pp$) system, just above the highest energy laboratory measurements of $\bar{p} p$ cross section and 2 orders of magnitude above the highest $pp$ measurements, but close to where one 
may argue that new physics 
such as supersymmetry, technicolor, or low scale gravity may 
begin to make contributions to $pp$ scattering.  Without adopting any particular symmetry or dynamics, one may thus speculate that some change in the fundamental hadron interactions might be the cause of the knee in the cosmic ray spectrum\cite{kn1,kn2, masip1, masip2, wigmans, nikolsky}.  
To explain the existence of the knee, one may propose that the incident spectrum obeys a single power law, but beyond a certain energy the collision between the cosmic ray 
primary and the target atmospheric nucleus is dominated by interactions that
lead to an enhanced cross section and production of unobservable, weakly interacting
particles, leading to missing energy. Hence the energy deduced from the shower does not correctly estimate
the energy of the primary, leading to the knee in the spectrum.

In the present paper we pursue the consequences of this hypothesis. In particular
we are interested in determining whether the measured nucleon-nucleon
cross section at ultra high energy using cosmic ray data \cite{bal,honda}
is consistent with the cross section required by this mechanism. Furthermore we 
are interested in determining how the new physics explanation of the cosmic
ray knee may be tested by  
observations at the Large Hadron Collider \cite{totem}.
Analysis of the impact of changes required in the nucleon-nucleon cross section
within this framework, analysis necessary for comparison both cosmic ray and 
collider data, is so far not available in the literature. We present here a step toward filling this gap.

We first assume that the 
cross section for new particle production adds
incoherently to the standard QCD nucleon-nucleon cross section, with a large portion that does not contribute to the observable shower. 
The new particle production process turns on only at very high energy
and hence this would predict a sudden increase in 
the effective nucleon-nucleon
cross section beyond the threshold for new physics.
This simple model for the cross section, however, may not capture all
the essential physics of the process. At high energy the nucleon-nucleon 
cross section becomes sufficiently large so that it may be more appropriate
to apply a black disc model to compute the proton-air nucleus cross section,
relevant for cosmic ray collisions. 
In this case the proton-air nucleus total cross section
has very weak dependence on the $pp$ cross section. The scattering
involves multiple interactions of the incident proton with the nucleons
in the nucleus, which itself involve multiple interactions of partons. 
If there exists a new physics interaction with an interaction strength
comparable to the QCD interaction at energy of the order of a GeV, 
it is very likely that 
almost every proton-nucleus collision involves at least one new physics
interaction. This will imply a considerable modification of the first approach
outlined above.   

An additional complication is that the nucleon-nucleon cross section
is dominated by soft hadronic physics and hence cannot be computed
from first principles. Although a precise first principle calculation 
of the cross section may not really be required for a reliable analysis of the cosmic ray
data, one does require a good understanding of the particle production 
rate in order to accurately extract the energy of the primary particle 
from observations. In the absence of a first principles calculation,
one must extrapolate a model calculation to high energies. 
Such an extrapolation may lead to large systematic errors
in the extraction of basic parameters such as the energy of the primary
particles.  

We adopt several simple models for the new physics cross section and explore a wide
range of parameter values in each. These cross section models are given
in Section 2. We use the recently published, high statistics data 
for the cosmic ray spectrum from the Tibet III experiment \cite{tibet} to implement our program. Our fits to the cosmic ray spectrum in the knee region  
 and the resulting prediction of the $pp$ cross sections are discussed in 
Section 3. The new physics scale and the fraction of missing energy must be such that the Tibet data are well fit and at the same time, reasonable agreement with the published ultra high energy cross section data \cite{bal,honda} is 
obtained. 
 In every case, we find parameter sets that give a good description of the Tibet III spectrum with a single power law injection spectrum.
However the required cross section values are found to be much 
larger in comparison to the values extracted from cosmic ray data. This
is discussed in Section 4. In Section 5 we discuss the black disc limit
of the cross section.  Here the multi-particle scattering obscures the link between the p-air nucleus total cross section and the $pp$ total cross section.
 In this limit it is not possible to write the total cross section as the sum
of the contribution due to standard model and that due to new physics. 
We make a fit to the cosmic ray flux by extrapolating our fit to the accelerator $pp$ total cross section to the energy range of the Tibet III data, modeling the fraction of
energy loss with a power law dependence on energy.  We find a good description of the Tibet III data with this simple, alternative picture as well.
Finally we conclude in Section 6.

\section{The total pp cross section: old and new physics}
We assume that the total proton-proton cross section derives contribution from both known physics (Standard Model, laboratory data) as well as some new physics effects. We express the total pp-cross section, $\sigma_{tot}$ as,
\begin{equation}
\sigma_{tot} = \sigma_{sm} + \sigma_{np}
\label{eq:sigmatot}
\end{equation}
where $\sigma_{sm}$ is the pp-cross section calculated using known physics while $\sigma_{np}$ is the cross section due to new physics. 

This model is analogous to those used in early studies of the growth of the $pp$ and $p\overline{p}$ cross sections with energy \cite{chl,ek,halz,ac-lvz,hh,gh}.  There the cross section was represented as $\sigma_{tot} = \sigma_{soft}+\sigma_{QCD}$, where $\sigma_{soft}$ incorporates the long distance, non-perturbative component and $\sigma_{QCD}$ incorporates the increasing number of perturbatively calculable jets. This early picture treats the ``soft" part as a constant, fixed at energies just above the resonance region, and ascribes the growth to the sum of perturbative QCD ``mini-jets".  The picture, though simple and intuitively appealing, basically the one adopted in \cite{kn1},  runs into contradiction with partial wave unitarity \cite{rd}, which can be addressed by adopting a diffractive approach, implemented by an impact-parameter representation \cite{dp, dp2}; later studies incorporated unitarity constraints \cite{bhm, bhs1, bhs2,bh1,bh,block}\footnote{Refs. 22 - 26 incorporate $pp$ and $p\bar p$ data as low as 6 GeV and also discuss simultaneous fits to the cosmic ray cross section data.}.  We will explore two models here, where the ``soft" term will be replaced by several parameterizations of the ``standard model" part of the high energy cross section and the ``mini-jet" portion of the cross section by models that produce a rapid rise of the cross section in the PeV region.

If the new physics process involves production of new, exotic particles that interact weakly with the atmospheric environment,  they will not contribute to shower energy. 
For incoming cosmic rays with energies above the new physics scale, the observed shower energies will then be lower than the energies of the cosmic rays that initiated them.  There would be a depletion of events above this scale, and a corresponding enhancement below. There would be an apparent ``knee" in the cosmic ray spectrum.  

In Eq. \ref{eq:sigmatot} we display our working model for
$\sigma_{tot}$. This model assumes that one can cleanly separate the events
in which either only the standard model interactions contribute from 
those where only the new interactions contribute. This leads to a convenient probabilistic 
approach that we outline next.

At high energies a cosmic ray proton primary with energy E interacts with the atmospheric
nuclei with probability

\begin{equation}
 P_{sm}(E) = \sigma_{sm}(E)/(\sigma_{sm}(E) + \sigma_{np}(E)) 
\label{eq:Psm}
\end{equation}
through the standard channel and with probability 

\begin{equation}
P_{np}(E) = \sigma_{np}(E)/(\sigma_{sm}(E) + \sigma_{np}(E))
\label{eq:Pnp}
\end{equation}
through the new channel.
Here the $\sigma_{sm}$ and $\sigma_{np}$ refer to the cross sections of
primary cosmic ray particles with air nuclei. However we shall assume that
we can
replace them simply by the corresponding values for the $pp$ cross section. 
This is justified as long as $\sigma_{sm}$ and $\sigma_{np}$ scale 
proportionately as we go from $pp$ to $p$-nucleus or nucleus-nucleus 
scattering for light nuclei. The $p$-nucleus cross 
section can be computed using the Glauber multiple diffraction theory
\cite{Glauber,Gaisser,bhs2}. One finds that the $p$-air cross section,
computed in Ref. \cite{bhs2}, depends almost linearly on the $pp$
 cross section as long as the cross section is not very large, 
suggesting that it may be reasonable to apply the 
Glauber theory to the two cross section $\sigma_{sm}$ and $\sigma_{np}$ separately  
and to adopt the scaling proportionality.
However, as discussed in the introduction, this assumption may break down
at ultra high energies, where the cross sections become very large.

We next assume that if the cosmic ray primary interacts through the new channel
then total energy detected is $E=yE'$, where $E'$ is the total energy of the incident cosmic ray and $y \leq 1$ is an energy loss factor that characterizes the new physics, assumed here to vary slowly enough to be effectively energy independent.  Because proton collision can be initiated by either the new physics or the standard model interaction, we find it useful for later application to introduce an effective, energy dependent $y_{eff}$

\begin{equation}
y_{eff}(E) = P_{sm}(E) + y P_{np}(E),
\label{eq:yeff}
\end{equation}
where $y$ is the value that follows from the fit of the model  data, and $y = 1$ for the Standard Model.  Clearly for $E \leq E_{np}$, where $E_{np}$ is the threshold scale of new physics, $y_{eff}=1$.

Next let $\phi(E)= {\cal N}E^{-\gamma} \exp (-E/E_{cut})$ be the incident galactic 
cosmic ray flux, with $\gamma\approx 2.7$ and the cutoff $E_{cut} \sim 10^8$ to $10^9$ GeV, representing the end of galactic sources to the CR spectrum. Here ${\cal N}$ is an overall normalization. 
The observed intensity at energy $E$ is then given by the incident intensity at $E$ times the probability that the observed showers are initiated by a standard physics collision at that energy plus the incident intensity at energy $E/y$ times the probability that the observed showers are initiated by the new physics interaction at the energy $E/y$, which takes into account the energy loss due
to new physics. We can express the {\it observed} differential flux spectrum, $\phi_{obs}(E)$, as

\begin{equation}
\phi_{obs}(E) = \phi(E) P_{sm} + \int_{E}^{\infty} dE' \delta (E-y E') 
\phi(E') P_{np}(E')   . 
\end{equation}

Pulling out an overall factor of the incident intensity at $E$,  we write the {\it observed} flux at $E$ as

\begin{equation}
\phi_{obs}(E) = {\cal N}E^{-\gamma} \exp (-E/E_{cut})\left[P_{sm}(E)+y^{\gamma-1}e^{-(E/yE_{cut})(1-y)}P_{np}(E/y)\right]. 
\end{equation}
With this general framework, we are equiped to determine the new physics parameters required to fit the  data for the cosmic ray spectrum covering the knee region.  
To proceed, we need to specify the laboratory-measured $pp$ and $p\overline{p}$
total cross sections and the generic, new-physics models
we will use for our study.

\subsection{The laboratory-measured, ``standard model" pp cross sections}

Let us first fit the proton-proton cross section data above 50 GeV \cite{pdg08} from accelerator measurements, but  excluding the high energy points reported by the Fly's Eye \cite{bal} and AGASA \cite{honda}, which lie in the energy range 
 $2\times 10^7$ GeV $\leq E \leq  4.8\times 10^8$ GeV, an order of magnitude and more above the energy where the knee in the spectrum appears at $2 - 4\times 10^6$ GeV.  We can then ask whether the extrapolation of the fit to the laboratory cross section measurements plus the new physics cross section found from adjusting its parameters to fit the CR spectrum with an assumed flux that falls with a fixed power law, namely $E^{- 2.66}$ below the knee, according to the Tibet III data, reproduces the direct CR cross section measurements lying above the knee region of the spectrum. 

In order to have a reasonable fit to the laboratory energy data \cite{pdg08}, we assume
the following ``Froissart bound saturation" form for $\sigma_{sm}(s)$ \cite{bh1,bh,block}, known to yield good fits to the laboratory data above about 10 GeV.  The fit function is
\begin{equation}
\sigma_{sm} = C_0 + C_1 \log(s) + C_2\log^2(s),
\end{equation}
where $s = 2 m_p^2 + 2 m_p E$ is the center of momentum frame total proton-proton energy squared.
Since we are strictly interested in the high energy regime and a fit to the cosmic ray cross section data we fit only the laboratory data set above 50 GeV.
The resulting parameter values are given in Table \ref{tab1}.
As an additional exercise, we include three $p \overline{p}$ points from collider measurements, which lie several orders of magnitude above the available $pp$ laboratory values.  Including these points that extend nearly to the energy corresponding to that of the knee, we find the parameter values given in the second
row of Table \ref{tab1}.

\begin{table}
\begin{tabular}{|lllll|c|}  \hline
Data set    & $C_0$ & $C_1$ & $C_2$ & $\chi^2$ & d.o.f.   \\ \hline
$pp$ $\geq$ 50 GeV &  46.60 & $-6.875$ & 1.472 & 10.5 & 29        \\ \hline
$pp$ $\geq$ 50 GeV +$p\overline{p}$ & 41.19 & $-3.863$ & 1.077 & 15.6 & 32       \\ \hline
\end{tabular}
\caption{First row: parameters, $\chi^2$ and number of degrees of freedom for $Log^2(E/E_0)$ fit, Eq. 7, to fixed target $pp$ total cross section data with E $\geq$ 50 GeV. Second row: same data plus 3 collider points.}
\label{tab1}
\end{table}

\begin{figure}[htb]
\includegraphics[height=2.5in,width=3.5in,angle =0]{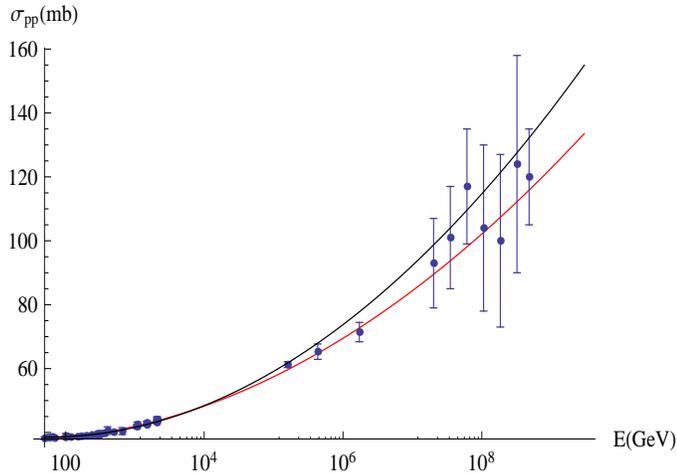}
\caption{$pp$ cross section above 50 GeV plotted against energy showing fits (see text) that include three collider $p\overline{p}$ data points, lower curve, and do not include them, upper curve. The cosmic ray data are displayed, but are not included in the fit.
} 
\label{fig:HEfit}
\end{figure}

The plots of the cross section fits versus data are shown in Fig. \ref{fig:HEfit}, which includes the highest energy points from the cosmic ray data, not included in the fit.

\subsection{Two new physics cross section models}

We start our discussion with a generic new physics cross section, which 
asymptotes to a constant times the extrapolated standard model cross section $\sigma_{sm}(E)$.  We chose a simple parametrization that may be expressed as,

\begin{equation}
\sigma_{np}(E) = \alpha \sigma_{sm}(E) \tau(E) g(\tau(E)) \nonumber
\end{equation}
Here $\alpha$ is a dimensionless constant, $\tau(E) = M_0^2/s$, $s \simeq 2 m_p E $ and $M_0$ is
the scale of new physics. In the nucleon-nucleon collision, the factor $\tau(E) g(\tau(E))$ represents the differential gluon-gluon luminosity having invariant mass-squared $\hat{s} \geq M_0^2$ \cite{ehlq}. 
The function $g(\tau(E))$ takes the form 
\begin{equation}
 g(\tau) = \int^1_{\tau} dx f(x) f(\tau/x), 
\end{equation}
The function
$f(x)$ is parametrized as \cite{bfhmv}
\begin{equation}
f(x) =  \sqrt{n+1}\frac{(1-x)^n}{x}.
\end{equation}
In the limit $E/(M_0^2/2m_p) \rightarrow \infty$, $\tau \rightarrow 0$
but $\tau g(\tau)\rightarrow 1$, so $\tau g(\tau)$ acts as a ``step" function that rises from 0 at energies below the new physics threshold at $E =M_0^2/(2 m_p)$ to 1 at higher energies. In this limit,
$\sigma_{np}(E)/\sigma_{sm}(E) \rightarrow \alpha$; for this model the total cross section approaches  $(1+\alpha)\sigma_{sm}$ at energies much larger than $M_0^2/(2 m_p)$.\footnote{This is the same $log^2(E)$ asymptotic behavior as the ``Froissart bound" fit to the data \cite{bh}, but with a strength 1+$\alpha$ times as great.} 

Our second model is a generic parameterization that copies the $\log(E)$ plus $\log^2(E)$ fit that is very successful at describing the laboratory total $pp$ cross section, as we see explicitly in the following section. Introducing a new scale $E_n$, we assume the new physics total cross section to be of the form
 
 \begin{equation}
 \sigma_{np}(E) = D_1 \log(E/E_n)+D_2 \log^2 (E/E_n),
 \end{equation}
with $\sigma_{np}(E)$ = 0 if $E \leq E_n$, and we fit the flux to the form of Eq. 6. The fit parameters are $E_n$, the new physics scale, $y$, the energy loss parameter, 
$\mathcal {N}$, the normalization, $ D_1$ and $D_2$, the coefficients of the linear and quadratic $\log(E/E_n)$ terms in the new physics cross section.

\section{New physics models fit to Tibet III data}

First we explore the parameters of the model by fitting it to the average of the Tibet III QGSJET-HD \cite{Kaidalov,Kalmykov} and SIBYLL \cite{Fletcher} data in the knee region, using the difference of each from the average to estimate a systematic error.  The total errors are then taken to be the rms value of this systematic error estimate and the quoted, essentially common, statistical errors associated with the two analysis packages.

\begin{figure}[htb]
\includegraphics[height=2.5in,width=3.5in,angle = 0]{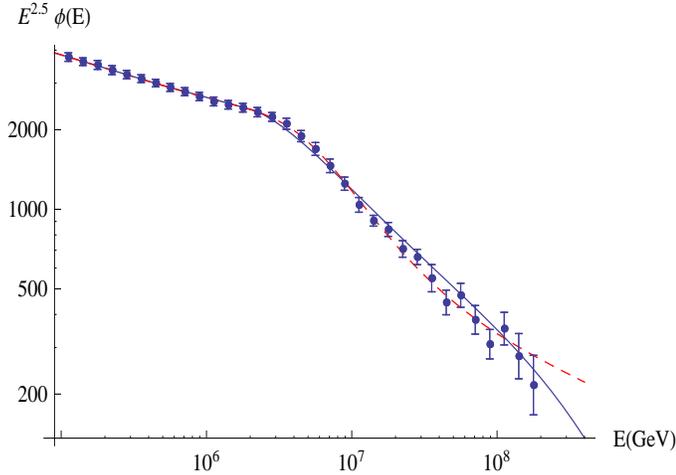}
\caption{Energy spectrum with $E$ along the $x$-axis and $E^{2.5}\times flux$ along $y$.  The dashed curve is the model 1 fit and the solid curve is the model 2 fit, using only pp laboratory data as input, with no constraints from the cosmic ray cross section estimates. } 
\label{fig:spectrum_fit1}
\end{figure}

We compare model 1 with the data by fitting the form

\begin{equation}
\phi_{obs}(E) = {\cal N}E^{-\gamma} \exp (-E/E_{cut})\left[{1\over 1+\alpha\tau g(\tau)} + {\alpha y^{\gamma-1}e^{-E/yE_{cut} + E/E_{cut}}\over \alpha + 1/[(\tau y) g(\tau y)]}\right] 
\end{equation}
to the data,  finding values for parameters $\cal{N}$, $\alpha$, $M_0$, $y$, and $E_{cut}$.
Because of our definition of model 1, $\sigma_{sm}$ does not appear in $\phi_{obs}$. An example fit and comparison of the model with the data 
is shown in the dashed curve in Fig. \ref{fig:spectrum_fit1}. Here, following \cite{kn1} for illustration,
we set the parameters $n=6$ and we set $\gamma$ = 2.666, the value we obtain for the region $E \leq 10^6$ GeV.  

The fit is good, creating the features of the data with a uniformly falling power law spectrum cut off at $10^{10}$ GeV, with a new physics cross section that is proportional to the SM cross section with an energy dependent coefficient added to the SM cross section.  The fit finds a new  physics characteristic scale just below 900 GeV and the energy loss parameter of $y \simeq$ 0.2, which ensures that most of the energy from a cosmic ray collision is invisible.

Next we also fit the data using 

\begin{equation}
\phi_{obs}(E) = {\cal N}E^{-\gamma} \exp (-E/E_{cut})\left[{1\over 1+R(E)} + {R(E/y) y^{\gamma-1}e^{-E/yE_{cut} + E/E_{cut}}\over 1 + R(E/y)}\right], 
\end{equation}
where $R(E) = [D_1 \log(E/E_n) + D_2 \log^2(E/E_n)]/[C_0 + C_1 \log(s) + C_2 \log^2(s)]$ and $s = 2 m_n^2 + 2 m_n E$, 
that follows from model 2. In Fig. \ref{fig:spectrum_fit1}, the solid curve shows the case where only the $pp$ data are used for the laboratory physics input, which are fit separately with the form $\sigma_{pp}^{SM}(E) = C_0 + C_1 \log(s) + C_2\log^2(s)$, as described in section 2.1.   When we include collider data points for $p\overline{p}$ \cite{pdg08} in the SM fit, the result differs little from the result for only $pp$ data shown in Fig. \ref{fig:spectrum_fit1}.  For the fit without the collider points, the essential parameter values are $E_n$ = $2.17\times 10^6$ GeV and $y$= 0.20, while for the fit with the these points, we find  $E_n$ = $3.04\times 10^6$ GeV and $y$ = 0.20.  Again the fits are quite good.

\section{ Fitting the all proton-proton cross section data and Tibet III data with new physics}

Referring to Fig. \ref{fig:HEfit}, we see a gap between the highest energy pp lab data and the lowest energy CR data, a gap which contains the energy range of the knee in the CR spectrum.  The question we address next is: ``What is the behavior of the total $pp$ cross section, new physics plus measured lab values, implied by our new physics analysis of the Tibet III spectrum, which spans the energy range from $E = 1.12\times 10^5$ to $E = 1.78\times 10^8$?"  This range includes the collider cross section measurements and most of the CR cross section measurements.

\begin{figure}[htb]
\includegraphics[height=2.5in,width=3.5in,angle =0]{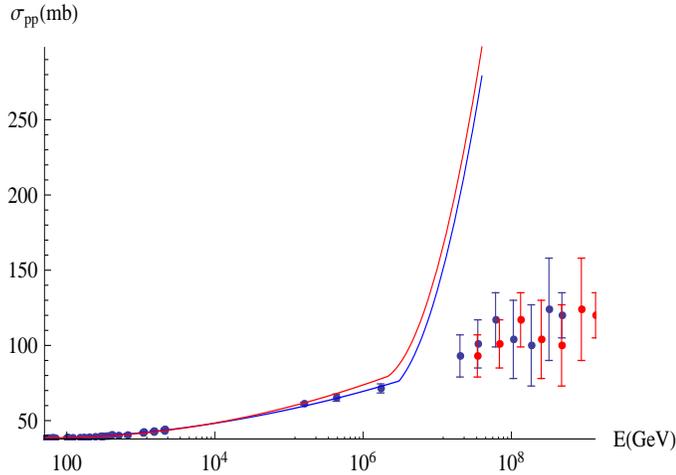}
\caption{$pp$ cross section data above 50 GeV compared to model 2 cross sections fitted to laboratory data with 3 collider $p\overline{p}$ data points, lower curve, and without these points, upper curve.  The new physics parameters are determined by fitting model 2 to the Tibet III flux data. The high energy data points show the rescaled (see text) cross sections extracted from the cosmic ray data. The
red and blue points employ rescaling corresponding the upper and lower curve 
respectively.} 
\label{fig:totfitvsdata2}
\end{figure}

In order to compare the cross sections extracted from cosmic ray data to
our predicted values, including new physics contributions, we must
appropriately rescale the energy assigned to these cross sections values.
This is because the analyses \cite {bal,honda} that led to the highest energy cross section points in Fig. \ref{fig:totfitvsdata2}  assign primary, collision energies to the events based on standard analysis of totally inelastic p-air collisions that produce showers containing {\it all} of the incoming primary cosmic ray energy.  If instead, there is substantial energy missing, energy in a form not accessible to the detectors, then the primary energy must be higher than contained in the observed shower.  This implies that the energy assigned to events is too 
low by the factor $1/y_{eff}$, where $y_{eff}$ is the effective energy loss
parameter as defined in Eq. \ref{eq:yeff}. Therefore the energy of each cosmic ray cross section point must be rescaled by this factor. 

In Fig. \ref{fig:totfitvsdata2}, we show the total cross sections, new physics plus extrapolated laboratory fits, plotted versus energy for new physics model 2.  This is the case where the fitting function is linear plus quadratic in Log(E/$E_n$), as described in the preceding section.  We show both of the cases we considered, the $pp$ lab data only and the $pp$ lab data plus three $p\overline{p}$ points.  The details differ, but in both cases the cross section required to produce the apparent knee in the CR spectrum rises explosively at energies below those where cosmic ray measurements constrain the cross section, albeit with large uncertainties.  The model 1 case is quite similar to that shown for model 2 in Fig. \ref{fig:totfitvsdata2}.

We find that the predicted cross sections show dramatic disagreement with 
those measured from cosmic ray data. Hence the simple cross section model,
Eq. \ref{eq:sigmatot}, is ruled out by the cosmic ray cross section data.  
However as discussed in the introduction, when the cross section values become
so large, we expect that a typical $pp$
collision involves multiple scatterings of the elementary quarks and gluons.
Hence once the energy becomes sufficiently high 
a typical process might involve many standard model interactions
at the parton level and perhaps one (or few) new physics interactions.  
In this case there may be no clean separation between the standard model
and new physics contributions to the $pp$ cross section. 
This distinction gets further clouded when we consider the 
proton-nucleus collisions, relevant for cosmic ray showers. We take up these questions in the next section.

\section{Black Disc Limit} 
At ultra high energies the simple formula, Eq. \ref{eq:sigmatot}, employed 
so far for $pp$ scattering cross section may no longer be applicable. As
discussed in Section 1, in this limit it may not be possible to cleanly separate
the contributions entirely due to standard model interactions from 
those due to new physics interactions. Lets assume that the new physics
interaction strength is comparable to that of the QCD coupling at the energy scale
of a few GeV. In this case a typical $pp$ collision may involve
many standard model interactions along with some new physics interactions. 
This is even more true in the case of proton-nucleus collisions relevant
for cosmic rays. In this case, as the cross section grows the incident proton undergoes multiple scatterings
with the target nucleons such that the proton-nucleus cross section
becomes nearly independent of the underlying $pp$ cross section above some black disc threshold \cite{bhgny,Gaisser}. 
Thus once 
we cross the threshold for production of the ``invisible" particles 
associated with new physics it may be more natural to assume that 
almost every collision involves a certain amount of energy loss due
to new physics interaction. The amount of energy loss per collision would
increase with energy due to increase in probability for new physics
to contribute.  

\begin{figure}[htb]
%\begin{figure}[htbp]
%\begin{center}
%\epsfile{file=
\includegraphics[height=2.5in,width=3.5in,angle =0]{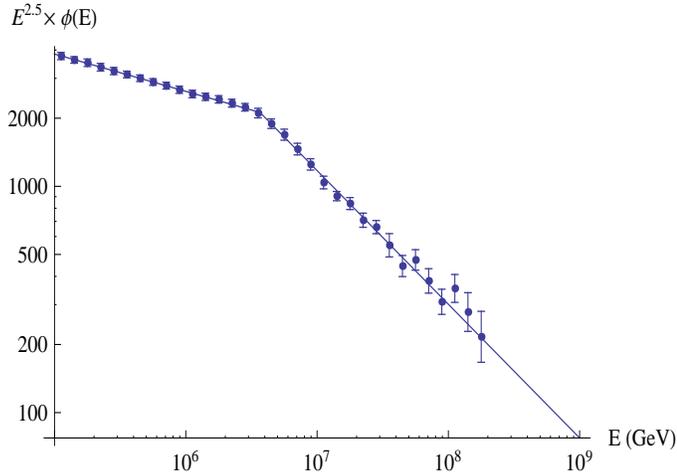}
%,scale=0.8}
%\caption{{\bf default}}
%\label{default}
%\end{center}
%\end{figure}
\caption{The Tibet III data is fitted with a cross section model that is simply an extrapolation of the lower energy fit up through the $p\overline{p}$ collider data.  At a threshold $E_T$, new physics sets in that produces a fraction of missing energy that is controlled by a power  $p$.  The threshold is at $3.68\times 10^6$, and the $p$ value is $-0.24$, in the curve through the data shown in the figure.}
\label{fig:y-powerlaw}
\end{figure}

To implement the idea just described, we assume that the process is 
described by a single total cross section $\sigma(E)$, which has a threshold 
for production of new invisible particles and, in general, a change 
in magnitude. The energy loss parameter, $y$, is taken to be a non-trivial function of energy, $f(E)$, rather than a constant as in the earlier discussion.  We will take it to be a power of energy shortly, but consider it to be unspecified for the moment. The number of events per unit of energy, time, solid angle and area, suppressing acceptance,
can be written

\begin{eqnarray}
\phi_{obs}(E)=\int_{E}^{\infty} dE'   \Big[\theta(E_T-E')\delta(E-E') \phi(E')\nonumber\\
 \ \ \ \ + \theta(E'-E_T)\delta(E-f(E'))\phi(E')\Big],
\end{eqnarray}
where $E_T$ is the threshold for onset of new physics effects. Performing the trivial integration, we find

\begin{eqnarray}
\phi(E)_{obs}=\phi(E)\theta(E_T-E)+ 
 {\phi(E_0(E))\over df(E_0(E))/dE}\theta(E_0(E)-E_T),
\end{eqnarray}
where $E_0$ is the solution to $E=f(E')$.  Assuming a power law form $f(E)=E^{1+p}/E_T^p$, we find a solution for the power needed to produce the break at the knee in the Tibet III spectrum with $p=-0.24$, and $E_T =3.68\times 10^6$ GeV. 
Within the uncertainties in these extrapolations \cite{luis}, there is room for a new physics contribution to the $pp$ cross section of the order of a factor two or so. If it has an energy -dependent missing energy fraction that grows with energy as modeled here, it can plausibly produce the observed knee in the cosmic ray spectrum, as illustrated for the case of a simple power law dependence, in 
Fig.\ref{fig:y-powerlaw}.

These results show that the situation changes when the cross section is saturating the  multiple scattering limit, where the cross section does not cleanly separate into several pieces at each energy, and the new physics fraction of visible energy increases with energy. In this case we are unable to predict the
dependence of cross section on energy. For example, we could change the cross section, while at the same time modifying the form of the energy dependence of the missing energy.  Hence we find that we are unable to test the new physics hypothesis
to explain the cosmic ray knee simply by total cross section measurement at the LHC. 
A more detailed analysis which involves studying the interaction strengths
of new physics processes is required.  First, a systematic failure to match the total final state energy with the LHC collision energy would have to be established to support the picture we present here.  Then, taking the TOTEM experiment for definiteness \cite{totem},  the measurements of the numbers of elastic and inelastic events, $N_{el}$ and $N_{inel} $, along with $dN_{el}/d|t||_{|t|=0}$, would have to be corrected to account for the missing energy to find the total cross section via the optical theorem and to determine the luminosity.  Conversely, if the cross section and luminosity are determined by these measured quantities without correction, then if an independent luminosity measurement is inconsistent with that determined by $N_{el}$ and $N_{inel}$ it could be evidence for new interactions producing undetected energetic particles, i.e. missing energy.    Such an independent luminosity determination uses the precise measurement of the horizontal vertex distribution at the interaction point (IP5 of LHC), as mentioned in the conclusions of the second paper cited in \cite{totem}. In the end, if no such inconsistencies are found, then our proposed explanation of the knee effect is ruled out.  If evidence for missing energy beyond that expected in the standard model is found, then the cross section, its energy dependence and the energy dependence of the missing energy effect must be able to account for the knee.  Of course it may be only part of the story behind the knee phenomenon, the rest coming from new astrophysics.

 %The proposal would of course be ruled out if 
%e does not find any new particles which escape detection in the collider. 

\section{Summary and Conclusions}

We have employed several models, generic parameterizations, of new physics 
interactions to show that the knee in the cosmic ray spectrum in the region 
$2-4\times 10^6$ GeV can be described successfully with only a few parameters.
We first wrote the total cross section as a sum 
of the contributions due to standard model and those due to new physics.
In this simple picture, the observed showers are initiated either by the standard model interaction or by the new physics interaction, with weights given by the respective cross sections. We hypothesized that above the threshold for new physics a fraction $y$ of the primary energy is in a form that the arrays do not detect, but that the observable energy is in showers similar to those modeled by the experiments.  We fixed the spectral index of the incident flux at the value determined from the data below the knee. 
%This simple model is reasonable in the limit of weak 
%new physics interaction strength. In that case one expects that the
%cross section is dominated by QCD with a small contribution from 
%new physics which adds incoherently to the standard model cross section. 
%However 
We found that the total proton-proton cross sections implied by the 
parameters determined by fitting the cosmic ray knee rise extremely rapidly in the energy range just below the knee. The result is that the discrepancy between the cross sections required to fit the knee with a single power law injection flux below about $10^9$ GeV appear to be completely ruled out by the estimates of the cross section published by the Fly's Eye and AGASA collaborations in the 1980's and 1990's.  
%However,those cross section measurements rely on the assumption that the primary interactions are inelastic. 

However the final $pp$ cross section at ultra high energies 
is so large that the picture which leads
to the simple form for the total cross section, Eq. \ref{eq:sigmatot}, 
is called into question. 
In this case multiple scattering is likely to dominate the nuclear collision and subsequent shower formation.  In this limit, the nucleus becomes a black disc
and one loses the clean separation
between the new physics and standard model contributions and 
furthermore loses the direct link between 
the nucleon-nucleon and the nucleon-air cross section. Therefore it
is not possible to test the new physics proposal purely on the basis
of comparison with $pp$ total cross section data. One would require a
detailed analysis of the LHC data in order to determine if there exists a
new interaction of sufficiently large strength along with sufficient 
missing energy in the final state to account for the knee in the CR spectrum.  We pursued the consequences of this black disc picture for the Tibet III data. We modeled the new physics as invisible energy loss with a power law dependence on collision energy that sets in at a threshold energy $E_T$, adopting a total cross section value extrapolated from our fit to accelerator data. Within this picture we also obtained a good fit to the cosmic ray flux in the vicinity of the knee. 

We conclude that, based on the published ultra-high energy total proton-proton and proton-antiproton cross sections and a proper rescaling of the cosmic ray data, a consistent picture of the CR knee in terms of purely new physics effects with an invariant mass scale in the 1-3 TeV range emerges. Moreover, looking at the limit where high multiple scattering approaches a black disk picture, while the fraction of energy loss increases with energy, a natural solution to the knee in the CR spectrum again emerges.

\section{Acknowledgements} D.W.M. thanks the Indian Institute of Technology Kanpur Department of Physics for support and warm hospitality during the time this project was initiated.  D.W.M. receives support from the U.S. Department of Energy under grant No. DE-FG02-04ER41308.

\end{document}